\documentclass[preprint,12pt,5p,times,twocolumn]{elsarticle} 
\usepackage{epsfig}
\usepackage{setspace}

\journal{Solid State Communications}

\begin{document}
\begin{frontmatter}
\title{A study of SnS thin films and its suitability for photovoltaic
application based on the existence of persistent photocurrent}

\author[khalsa,sc]{Yashika Gupta}
\author[khalsa]{P.Arun \corref{cor1}\fnref{fn1}}
\ead{arunp92@physics.du.ac.in}

\cortext[cor1]{Corresponding author}
\fntext[fn1]{(T) +91 11 29258401 (F) +91 11 27666220}
\address[khalsa]{Material Science Research Lab., S.G.T.B. Khalsa College,\\ 
University of Delhi, Delhi 110007, INDIA}
\address[sc]{Department of Electronic Science, University of Delhi-South
Campus,\\ New Delhi 110021, INDIA}

\begin{abstract}
Tin Sulphide is a layered compound which retains its structure when
deposited as thin films by thermal evaporation. The films were found to have
oriented growth with the direction of orientation changing with film
thickness. The film's morphology was found to change with orientation. The
poor conductivity of the thicker samples made it difficult to make
photocunductivity characterisation. However, unlike reported the thinner 
samples showed photo-sensitivity to be independent of film thickness and 
grain size with a high presistent photocurrent. With their absorption, 
photosensitivty, optimum 
band-gap and traps within the band-gap giving the charge carriers a longer
life-time, thin samples of tin sulphide gives adequate scope designing
efficient photovoltaics. The refractive index was modeled using Sellmeir's
model, while most of the previous studies talk of Wemple-DiDomenico single
oscillator model or Cauchy's dispersion relation. The Sellmeir's fitting 
parameters are reported which can be of use in ellipsometric studies.  
\end{abstract}


\end{frontmatter}

\section{Introduction}
Search for an abundant, inorganic and non-toxic material with good
absorption coefficient at relatively low thicknesses in solar cell
applications has narrowed down to Tin Sulphide (SnS) \cite{gaoN, noguchi}. 
SnS has been successfully fabricated by chemical methods 
\cite{gao1}-\cite{nair}, spray pyrolysis \cite{reddy1}-\cite{guneri1}, 
sputtering \cite{suzuki}, thermal evaporation \cite{noguchi},\cite{nahass}-
\cite{ogah} and electron beam assisted evaporation \cite{tanu}. The band-gap 
of the films were reported to lay between
1.3-2.5~eV \cite{gaoN},\cite{fada}-\cite{aaron} depending on the 
fabrication technique. With band-gap laying in this range, SnS is
capable of utilizing the visiable range of solar radiation falling on the
earth's surface.  

Literature shows that the present efforts are directed towards
finding an appropriate second layer for the diode formation with SnS. The
best efficiency reported is 4.4\% with Zinc Oxysulphide, Zn(O,S), acting as 
the n-layer \cite{rggordon}. Annealing \cite{anneal} and/or sulphurization
\cite{sulphur} of the SnS layer has also been suggested to improve the SnS
solar cell's efficiency. However, directed fundamental characterization of
SnS films especially on presistant photoconductivity is absent. In fact a 
search of the literature only shows a couple of articles on SnS presistant 
photoconductivity
\cite{johnson,ECS}. Considering a long carrier lifetime (in relation to
photoconductivity) would assist in improving the photovoltaics 
preformance \cite{soci}, an indepth corelation of carrier lifetime with 
film parameters is a must. We have tried to address this in the present work.

The present work reports the fabrication of p-type SnS thin films via
thermal evaporation method on glass substrate maintained at room
temperature. Characterization of the films shows that the properties, both
electrical and optical, are strongly related to the orientation of 
the unit cell which inturn depends on the film thickness.

\section{Experimental}
Thin SnS films of varying thicknesses were grown at rates greater than 
2~nm/sec on microscopy glass substrate maintained at room temperature by 
thermal evaporation of SnS pellets at vacuums better than 
${\rm \approx 4\times 10^{-5}}$~Torr using a Hind High Vac (12A4D) coating 
unit. The starting material was SnS powder of 99.99\% purity
provided by Himedia (Mumbai). The thicknesses of the films were 
measured using Dektak surface profiler (150). Bruker D8 
X-ray Diffractometer with copper target (${\rm CuK\alpha}$ radiation, 
${\rm \lambda \sim 1.5406~\AA}$) and Transmission Electron Microscopy 
(Technai T30U Twin) was used to determine the structure/crystallinity of the 
samples. The surface morphology of the samples was compared using a 
Field Emission-Scanning Electron Microscope (FE-SEM FEI-Quanta 200F) at an 
accelerating voltage of 10~kV. The optical studies of the films were recorded 
using an UV-Vis Double Beam Spectrophotometer (Systronics 2202). Hot-spot probe 
method was used to determine the conductivity type of the films, which 
without exception showed that the as-grown films were p-type in nature.

The photoconductivity experiments were carried out at room temperature using 
a source meter SMU-01 of Marine (India). A tungsten lamp was used as our 
light source. Contacts were made using silver paste which is known
to give good ohmic contacts with SnS \cite{pram, sato}

\section{Results and Discussion}

\subsection{The Structural and Morphological Analysis}
%%%%%%%%%%%%%%%%%%%%%%%%%%%%%%%%%%%%%%%%%%%%%%%%%%%%%%%%%%%%%%%%%%
\begin{figure}
\begin{center}
\epsfig{file=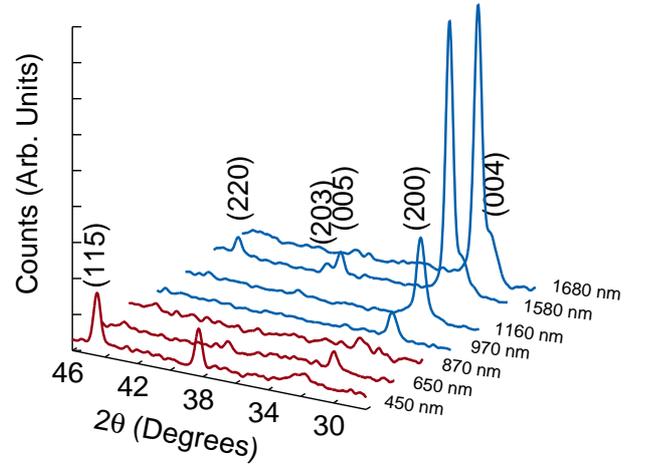, width=2.75in, angle=-90}
\end{center}
\caption{X-ray diffraction pattern}
\label{fig.1}
\end{figure}
%%%%%%%%%%%%%%%%%%%%%%%%%%%%%%%%%%%%%%%%%%%%%%%%%%%%%%%%%%%%%%%%%%

Fig~1 shows the X-Ray diffractograms of various samples whose thickness lie 
between 450 to 1680~nm. All the peak positions matched with those listed in 
ASTM Card No~79-2193 for SnS, having orthorhombic structure with lattice 
constants  ${\rm a \approx 5.673}$, ${\rm b \approx 5.75}$ and 
${\rm c \approx 11.76~\AA}$. As can be seen from the X-Ray diffractograms 
there is a change in orientation with increasing film thickness. Texture 
coefficient was estimated to determine the preffered orientation of 
crystallites in the as-deposited films using the formula
\begin{eqnarray}
T_{hkl}={\displaystyle\left[{I_{hkl} \over I_{o(hkl)}}\right] \over 
{1 \over N}\displaystyle\sum\limits_i^N \left[{I_{hkl} \over I_{o(hkl)}}\right]_i}\nonumber
\end{eqnarray}
%%%%%%%%%%%%%%%%%%%%%%%%%%%%%%%%%%%%%%%%%%%%%%%%%%%%%%%%%%%%%%%%%%%%
\begin{figure}[h!!!]
\begin{center}
\epsfig{file=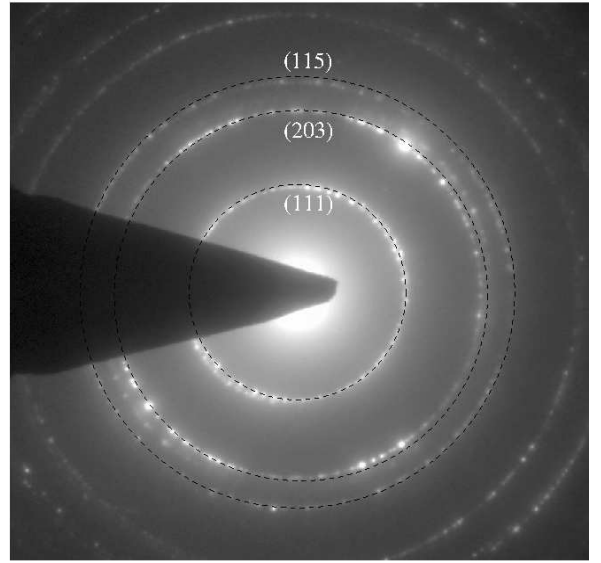, width=3in, angle=-0}
\end{center}
\vskip -0.5cm
\caption{SAED micrograph confirms the sample's crystallinity. Major rings
are marked and Miller indices indiated.}
\label{fig.22tem}
\end{figure}
%%%%%%%%%%%%%%%%%%%%%%%%%%%%%%%%%%%%%%%%%%%%%%%%%%%%%%%%%%%%%%%%%%

where ${\rm T_{hkl}}$ is the texture coefficient of (hkl) plane. This is
evaluated using the intensity of the (hkl) plane (${\rm I_{hkl}}$)
obtained experimentally from the X-ray diffractograms while ${\rm I_{o(hkl)}}$ is the 
intensity listed in the ASTM Card for the corresponding (hkl) peak and `N' is 
the total number of
diffraction peaks obtained in X-Ray diffraction. Preferential orientation 
(hkl) plane would have large texture coefficient compared to the other
planes. We obtained ${\rm T_{005}\approx 2.5}$ and ${\rm T_{200}\approx 3.35}$
for 450~nm and 1680~nm films respectively, showing a shift from (005)
orientation to (200) as film thickness increases. The Miller indices
suggests that the film's orientation moves from `xy' layers parallel to the
substrate (i.e. `c' or the long axis perpendicular to the substrate) to an
orientation where the `xy' layers are perpendicular to the substrate (long/
`c' axis parallel to the substrate). 
%%%%%%%%%%%%%%%%%%%%%%%%%%%%%%%%%%%%%%%%%%%%%%%%%%%%%%%%%%%%%%%%%%%%
\begin{figure}[h!!!]
\begin{center}
\epsfig{file=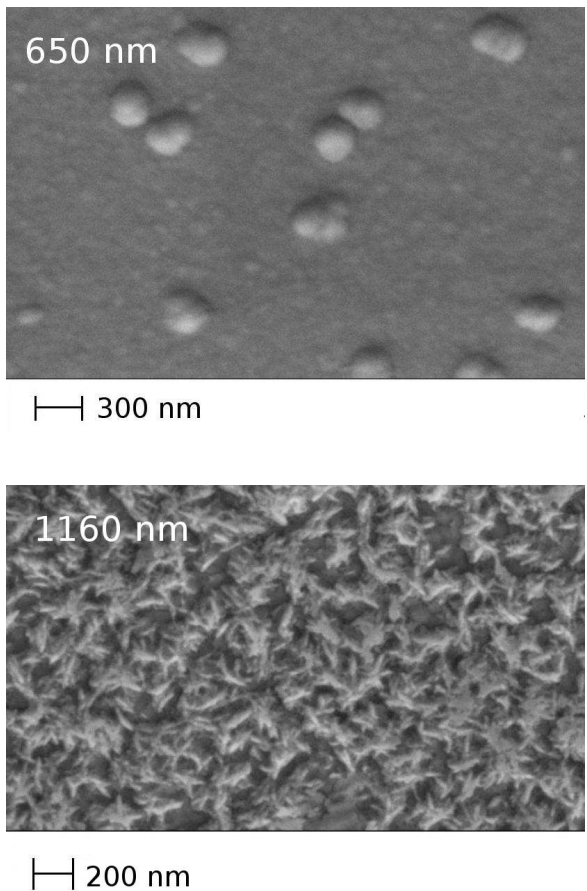, width=3in, angle=-0}
\end{center}
\vskip -0.5cm
\caption{SEM micrograph.}
\label{fig.22}
\end{figure}
%%%%%%%%%%%%%%%%%%%%%%%%%%%%%%%%%%%%%%%%%%%%%%%%%%%%%%%%%%%%%%%%%%

To confirm the crystalline nature of the grown films, we did a Selected Area 
Electron Diffraction (SAED) analysis of our 
650~nm sample using a Transmission Electron Microscope (Fig~2). The sharp
rings confirm the crystallinity of our samples. The inner most 
ring corresponds to (111) plane, while the second and 
third ring correspond to the (203) and (115) peaks of our X-ray diffractograms (Fig~1), the (111) peak did not show in XRD.
The SEM micrographs show two different morphologies, where thinner films showed 
spherical grains and thicker films exhibited `needle' shaped morphology 
(fig~3). The needle shaped morphology \cite{guneri}-\cite{park} is typical 
when the `c' or long axis is parallel to the substrate, thus confirming our 
assertion made while interpreting our X-Ray diffraction results.

The grain size of the SnS samples were determined from the XRD peaks 
according to the Scherrer’s formula \cite{cullity}
\begin{eqnarray}
D={ 0.9\lambda \over FWHM cos \theta}\nonumber
\end{eqnarray}
where `D' is the grain size, ${\rm \lambda}$ is the wavelength of the X-ray 
used (${\rm \lambda=1.5406\AA}$), FWHM is Full Width at Half Maximum 
intensity of the diffraction peaks and ${\rm \theta}$ is the Bragg's 
angle. Fig~4 shows the variation of grain size with film thckness.
%%%%%%%%%%%%%%%%%%%%%%%%%%%%%%%%%%%%%%%%%%%%%%%%%%%%%%%%%%%%%%%%%%%%
\begin{figure}[h!!!]
\begin{center}
\epsfig{file=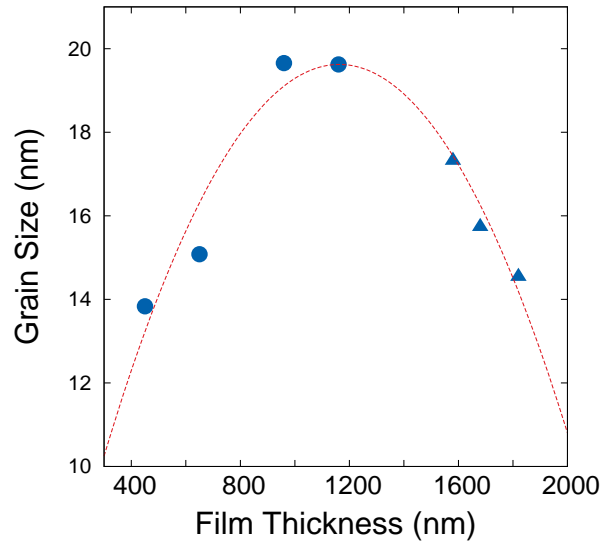, width=3in, angle=-90}
\end{center}
\vskip -0.5cm
\caption{Variation of grain size with thickness. Circles indicate grains with spherical morphology and triangles indicate needle-shaped grains.}
\label{fig.2}
\end{figure}
%%%%%%%%%%%%%%%%%%%%%%%%%%%%%%%%%%%%%%%%%%%%%%%%%%%%%%%%%%%%%%%%%%
Initially, the grain size increases with the thickness however, on reaching a 
maximum value it shows a decreasing trend with the film thickness.
Co-relating with our SEM micrographs, we find that it is at 
this ``point of inversion" thickness that the grain morphology changes from 
spherical to needle-shaped.

To study this variation in grain size with changing morphology, lattice 
parameters were calculated using the following formulae \cite{cullity}
\begin{equation}
{1\over d^2} = {h^2 \over a^2}+ {k^2 \over b^2}+{l^2 \over c^2}\label{eq.1}
\end{equation}
where a, b, c are the lattice parameters and (hkl) are the Miller indices 
reported in the ASTM card. It was observed that a residual stress acts along 
the `c-axis' of the orthorhombic SnS unit cell while the lattice parameters 
`a' and `b' were same as reported in the ASTM Card indicating no stress
along these directions. The strain along the `c' direction can be estimated
using
\begin{eqnarray}
\delta={c_{exp}-c_{ASTM} \over c_{ASTM}}
\end{eqnarray}
where `${\rm c_{exp}}$' is the experimentally obtained long lattice parameter  
and `${\rm c_{ASTM}}$' is that of a single crystal. 
%%%%%%%%%%%%%%%%%%%%%%%%%%%%%%%%%%%%%%%%%%%%%%%%%%%%%%%%%%%%%%%%%%%%
\begin{figure}
\begin{center}
\epsfig{file=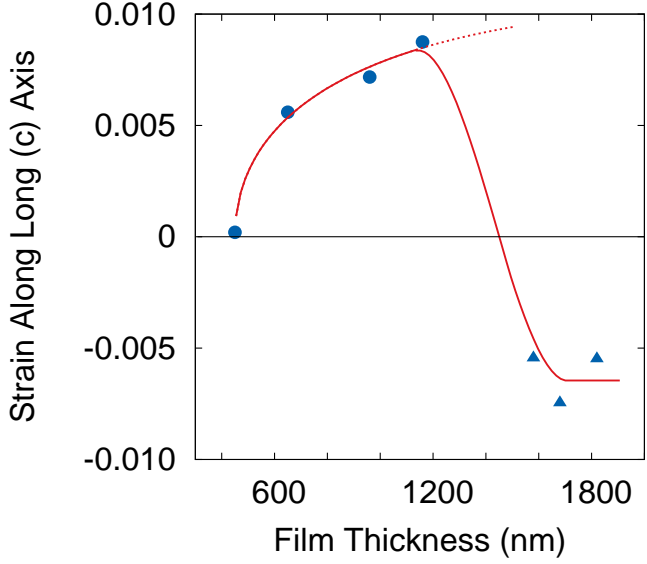, width=3in, angle=-90}
\end{center}
\caption{Variation of stress along the long (`c') axis with the film
thickness. Circles indicate grains with spherical morphology and triangles 
indicate needle-shaped grains. Lines are only visual aids to show trend in
variation.}
\label{fig.3}
\end{figure}
%%%%%%%%%%%%%%%%%%%%%%%%%%%%%%%%%%%%%%%%%%%%%%%%%%%%%%%%%%%%%%%%%%

The study of strain with film thickness shows an interesting trend (fig~5). 
Films with spherical grains show tensile strain (or tensile residual stress)
which increases with film thickness. Infact it shows a near parabolic trend
with film thickness, with rate of change shallowing as thickness increases
(lines in fig~5 are for visual aid). This may explain the increase in grain
size with film thickness. However, the films with the needle-shaped grains
show compressive stress acting on it. The magnitude of residual stress seem
to be constant with film thickness. The variation in residual stress with
film thickness would indicate that the nature of residual stress changes
with the long (`c') axis changing its orientation from perpendicular to the
substrate (tensile stress) to parallel to the substrate (compressive stress). 
This curve suggests that the change in orientation takes place in films of 
thicknesses greater than 1160~nm. 

\subsection{Photoconductivity}
%%%%%%%%%%%%%%%%%%%%%%%%%%%%%%%%%%%%
\begin{figure}[h!!!]
\begin{center}
\epsfig{file=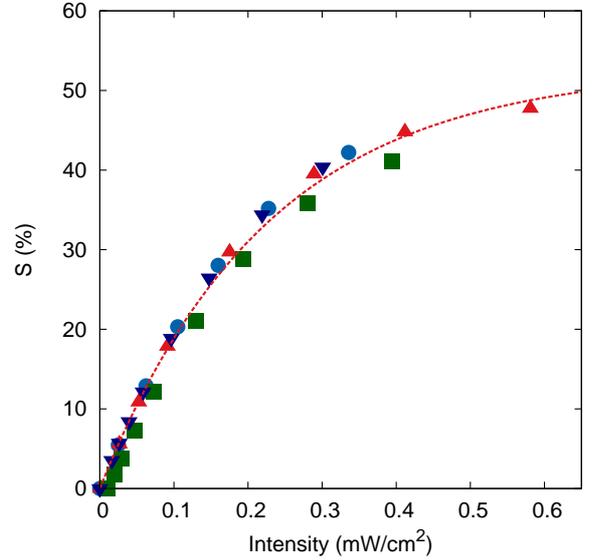, width=3in, angle=-90}
\end{center}
\vskip -0.6cm
\caption{Plot shows variation of ${\rm (\alpha h \nu)^2}$ with ${\rm h\nu}$
for two different film thicknesses (450 and 1820~nm). Extrapolating the best
fit line to the X-axis at y=0 gives the band gap of the respective films.}
\label{fig.sens}
\end{figure}
%%%%%%%%%%%%%%%%%%%%%%%%%%%%%%%%%%%%%%%%%%

To study the photoconductivity of the films, the \hbox{I-V} characteristics of the 
sample's measured with them exposed under a tungsten bulb.
Interestingly, the resistivity of the thicker films (${\rm d \geq 1160~nm}$) 
were found to be very high (beyond the measuring capability of the instrument).
Similiar observation was made by Wang et al \cite{gong}.
Considering that there is a sudden change in resistivity as the film
thickness is increased beyond 1160~nm, we believe this exceptionally high 
resistivity is related to the grain orientation (as seen in our structural 
analysis, thicker samples have `xy' plane perpendicular to the substrate). 
Sinsermsuksakul et al~\cite{gordan2}
have shown that the mobility of charge carriers is significantly lower along
the direction of Van der Waals forces. Hence, with electrodes made on the
film's surface, we can appreciate the high resistance in thick SnS samples.

``Sensitivity'' (S) is usually used as a figure of merit while discussing the 
photoconductivity of films and is given as \cite{bube}-\cite{kotadiya}
\begin{eqnarray}
S=\left(\sigma_L-\sigma_D \over \sigma_D\right)\label{s}
\end{eqnarray}
${\rm \sigma_L}$ and ${\rm \sigma_D}$ is the films conductivity measured
when exposed to light (L) and in dark (D) respectively. Sensitivity, hence is 
a measure of how much conduction increases in a sample with light intensity.
The increase in conductivity maybe due to an increase in charge carriers
\cite{petritz, woods} or
due to change in mobility \cite{slater}. From the charge carriers generated 
on illumination, a part would recombine 
with the respective opposite charges, thereby reducing the barrier potential
at grain boundaries. This inturn would lead to an increase in the charge
carrier's mobility. This is called grain boundary modulation \cite{kalita}. 
The photoconductivity is given as \cite{kotadiya}
\begin{eqnarray}
\Delta \sigma= q(\Delta p)\mu+qp(\Delta \mu)\nonumber
\end{eqnarray}
where `p' is the charge carrier concentration (here restricted to holes
considering p-type material is under discussion) and `q' charge associated
with single charge carrier, while ${\rm \mu}$ is the carrier's mobility.
Fig~\ref{fig.sens} shows the variation of sensitivity with intensity for four different
thicknesses (less than 1160~nm) of polycrystalline samples whose grain size were different. As
can be noticed, all the experimental data irrespective of the grain size
follows the same trend. This would suggest that the sensitivity of p-SnS is
independent of grain size, thus ruling out barrier modulation \cite{slater}. 
Thus, any
increase in conductivity is due to increase in photo-conductors or carriers
created by illumination. The above equation reduces to 
\begin{eqnarray}
\Delta \sigma= q(\Delta p)\mu\nonumber
\end{eqnarray}

Fig~\ref{fig.sens} is also indicative of the fact that sensitivity is
dependent on the illumination intensity. In fact it is functionally given as
~\cite{berg, johnson}
\begin{eqnarray}
S &\propto& I^\gamma \label{s2}
\end{eqnarray}
where `I' represents the illumination intensity. The exponential factor 
${\rm \gamma}$ can be evaluated from the slope of the straight 
line obtained when fig~\ref{fig.sens} is plot on a logarithmic scale. For our 
data we obtain ${\rm \gamma=0.876}$, where ${\rm \gamma < 1}$ indicates
existence of traps within the bandgap and also the presence of bimolecular 
recombination (recombination on band-gap transition) \cite{kalita,neetu}.
%%%%%%%%%%%%%%%%%%%%%%%%%%%%
\begin{figure}[h!!!]
\epsfig{file=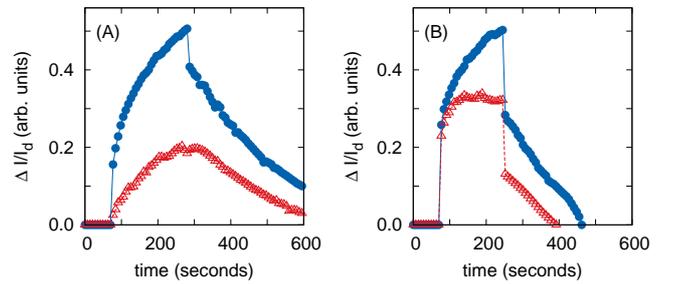, width=1.7in, angle=-90}
\vskip -0.4cm
\caption{Variation of photocurrent with time for films of various
thicknesses.}
\label{fig.ppc}
\end{figure}
%%%%%%%%%%%%%%%%%%%%%%%%%%%%%%%%%%

The existence of traps can also be appreciated from the rise and decay of
photo-current. Photo-current Decay (PCD) measurements of our
samples were made by applying a constant bias of 10V across them. Before
illumination, the samples were maintained in dark for ${\rm \approx 70~s}$. 
The photo-current was measured during illumination for ${\rm t>150~s}$,
followed by which we observed the current decay under the dark. Measurements
were made till the original dark current was achieved. Unlike sensitivity, 
two trends were obtained, for (a) thinner samples (${\rm d \leq 650~nm}$) and (b)
thicker samples (${\rm d \geq 870~nm}$).

The vertical rise and drop in current upon switching illumination ON and OFF
respectively is a result of charge carriers crossing the band-gap with no
contribution of traps. Whereas, the exponential rise and decay is trap
assisted photo-current. The trap assisted photo-current or the persistant 
photoconductivity (PPC) is dominant in thinner samples as compared to the 
thicker samples, as is evident by the larger vertical change regions of the
curve. The persistant current is given by the equation \cite{dang, li} as 
\begin{eqnarray}
I(t)=I_d + A_o exp\left[\left(-{t \over \tau}\right)^\beta\right]
\end{eqnarray}
where ${I_d}$ is the initial dark current, ${\rm \tau}$ the decay-time 
constant and '${\rm \beta}$' represents the stretching exponent constant 
(${\rm 0 < \beta < 1}$). The ${\rm \tau}$ values obtained by curve fitting 
are listed in Table~I. The large decay time indicates existence of trap
centers within the forbidden band-gap of the film \cite{ECS}. These trap 
centers increases the carrier life-time \cite{smith} in the as-grown p-SnS 
thin films. We find the charge carrier life-time systematically decreases
as film thickness increases. Considering an increased life-time implies a reduced 
recombination rate allowing more time for the carriers to move without
recombination, the material with long carrier life-time and good sensitivity 
would be a useful material for photovoltaics \cite{soci}.  
%%%%%%%%%%%%%%%%%%%%%%%%%%%%%%
\begin{table}[t]
{\bf Table I:} {\sl Decay time constants of various films obtained by curve
fitting eqn~(5) to photo-current data.}
\begin{center}
\begin{tabular}{cc}\hline
Thickness (nm) & ${\rm \tau}$ (seconds) \\
\hline
450 & 234.65 \\
650 & 207.47 \\
870 & 133.24 \\
970 &  90.30 \\
\hline
\end{tabular}
\end{center}
\end{table}
%%%%%%%%%%%%%%%%%%%%%%%%%%%%%%%%%%%%%%%%%%%%

\subsection{Optical Properties}
\subsubsection{Refractive Indices}
%%%%%%%%%%%%%%%%%%%%%%%%%%%%%%%%%%%%
\begin{figure}[h!!!]
\begin{center}
\epsfig{file=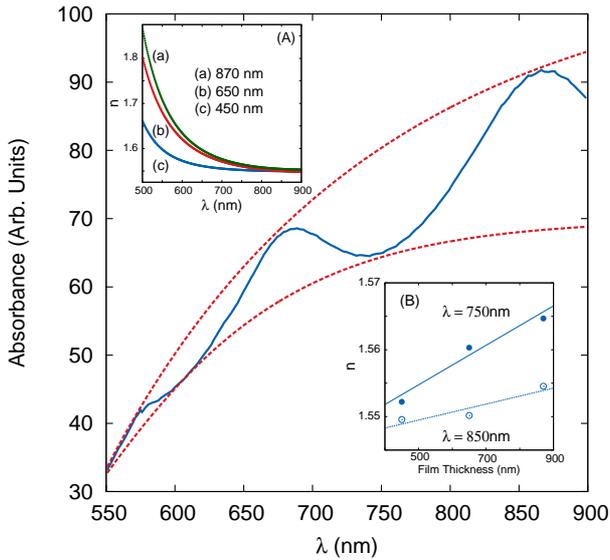, width=3in, angle=-90}
\end{center}
\vskip -0.6cm
\caption{Plot shows variation of ${\rm (\alpha h \nu)^2}$ with ${\rm h\nu}$
for two different film thicknesses (450 and 1820~nm). Extrapolating the best
fit line to the X-axis at y=0 gives the band gap of the respective films.}
\label{fig.refind}
\end{figure}
%%%%%%%%%%%%%%%%%%%%%%%%%%%%%%%%%%%%%%%%%%
The refractive indices of the films were evaluated for different wavelengths
using the standard Swanepoel's method \cite{swan1, swan2} from the transmission spectra of the
samples. Swanepoel's method involves drawing envolopes connecting the extrema
points of the the interference fringes appearing in the transmission spectra
(see fig~\ref{fig.refind}). Interestingly, these fringes did not appear in samples whose
thickness was more than 1160~nm. Again, like our observation made on
photoconductivity, films with thicknesses above 1160~nm coinciding with
structural orientation change, shows a marked change in behavior. 
In this study, due to lack of fringes in thicker samples and limitations
imposed by the technique, we report the variation of refractive indices seen
in just three of our samples. 

The inset (A) of fig~\ref{fig.refind} shows the variation of refractive
indices of the three samples with wavelength. The disperive trend of the
three films look similiar. In fact the trend follows the Sellmeir relation 
\cite{fuji} (curve fits have co-relation of 0.998) given below:
\begin{eqnarray}
n^2=A+\sum_j{B_j\lambda^2 \over \lambda^2-C_{oj}}\label{sell}
\end{eqnarray}
Sellmeir model pictures the solid to be made of dipole oscillators with `j' 
natural oscillation frequencies, given by ${\rm \omega_{oj}=2\pi c
/\lambda_{oj}}$ (in above equation ${\rm \omega_{oj}=2\pi
c/\sqrt{C_{oj}}}$). Interestingly, our results suggest SnS film's consist of 
single oscillators, all oscillating at frequencies corresponding to energies
within 2.72-2.83~eV. This is well above the sample's known band-gap (which we
shall discuss below), however the Sellimeir ``oscillators" do not absorb
away from the natural frequencies (our data is restrict for ${\rm \lambda
\geq 500~nm}$) and have zero broadening of the Lorentz absorption peak. This
implies that the refractive index is a real number with ${\rm
\epsilon_2=0}$. Table~II gives the coefficients of eqn~\ref{sell} that fit to
the experimental results which can be used for model fitting during
ellipsometric studies. An increasing `B' with film thickness suggests that 
the refractive index of the samples increases with film thickness for all 
wavelengths (see inset B of fig~\ref{fig.refind}). 

While our data clear shows that the refractive index fits Sellmeir's model,
previous works have reported SnS follows Cauchy's dispersion relation
\cite{botao} and Wemple-DiDomenico single oscillator model for refractive 
index \cite{wdd}.

\begin{table}[t]
{\bf Table II:} {\sl Comparison of properties of SnS based solar 
cells from the literature listed in order of increasing efficiency.}
\begin{center}
{\scriptsize {
\begin{tabular}{c c c c}
\hline 
  Film Thickness (nm) & A &  ${\rm B_1}$ & ${\rm C_{o1}}$ (${\rm \times 10^5}$)
${\rm nm^2}$\\ \hline
  450   & 2.278 & 0.0834 & 2.08\\
  650   & 1.983 & 0.3006 & 1.92\\
  870   & 1.963 & 0.3180 & 1.99\\
\hline
\end{tabular}
}}
\end{center} 
\end{table}

\subsubsection{Band-gap Variation}

%%%%%%%%%%%%%%%%%%%%%%%%%%%%%%%%%%%%
\begin{figure}[h!!!]
\begin{center}
\epsfig{file=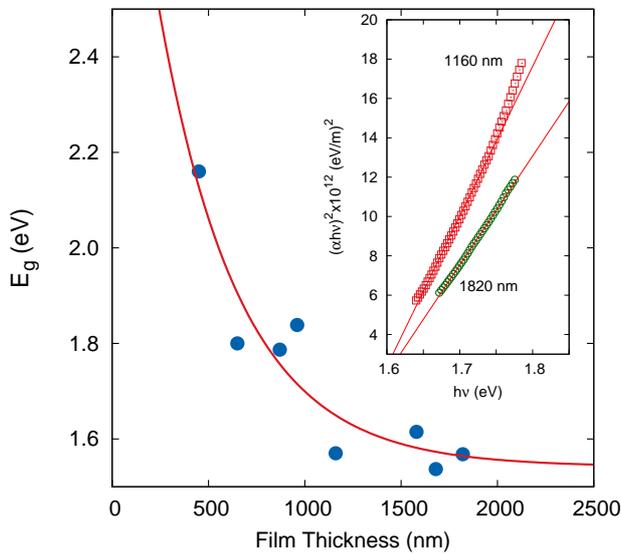, width=3in, angle=-90}
\end{center}
\vskip -0.6cm
\caption{Plot shows variation of ${\rm (\alpha h \nu)^2}$ with ${\rm h\nu}$
for two different film thicknesses (450 and 1820~nm). Extrapolating the best
fit line to the X-axis at y=0 gives the band gap of the respective films.}
\label{fig.eg}
\end{figure}
%%%%%%%%%%%%%%%%%%%%%%%%%%%%%%%%%%%%%%%%%%
The absorption spectra of the as-grown SnS films were studied in
the wavelength range 300-900~nm. The band gaps of the films were evaluated 
from the absorption spectrum using tha standard Tauc method \cite{tauc}. 
Although SnS is reported to have both direct and indirect
band-gaps~\cite{chen}, 
we obtained straight line fits only for plots between ${\rm (\alpha h\nu)^2}$ 
and ${\rm h\nu}$ (see inset of fig~\ref{fig.eg}), where`${\rm \alpha}$' is 
the absorption coefficient and ${\rm h\nu}$. This suggests that our p-SnS 
as-grown films have allowed direct band-gap~\cite{direct}. The band-gap shows 
an exponential decrease with film thickness attaining the value of bulk samples
(${\rm \approx 1.54~eV}$ \cite{xu}-\cite{basu}) for thicknesses greater than 1500~nm.

\section{Conclusions}
Tin Sulphide (SnS) thin films were fabricated by thermal evaporation on
glass substrates maintained at room temperature. The films were found to be
nano-crystalline with layered structure, whose layers were found to be
either parallel to the substrate or perpendicular to it depending on the
film thickness. The orientation was found to have an effect on the film's 
morphology and its properties, such as electrical conductivity. The poor
conductivity of thicker films caused by layers being perpendicular to the
substrate made it impossible to study the photoconductivity of these
samples. However, thinner samples showed good photo-sensitivity and we
observed presistant photoconductivity in them. This indicates the existence
of traps within the forbidden energy gap of thin samples. The charge
carrier's life time was found to decrease with increasing film thickness.
This would imply an improved photovoltaic performance in very thin samples
of SnS with charge carriers reaching their respective electrodes without
recombination. The optical studies showed a 2.2~eV band-gap for the thinner
SnS films with large refractive index following Sellmeir's dispersion
relation, ideally suited for simple photovoltaics
and plasmonic photovoltaics using SnS as the absorbing layer. 

\section*{Acknowledgemnt}
One of the authors (YG) would like to express her gratitude to DST~(India)
for the finanical assistance in terms of fellowship under the INSPIRE
program (Fellowship No. IF131164).

\end{document}